\documentclass[letterpaper, twocolumn, aps, prl, amsmath]{revtex4}

\usepackage{graphicx}
\usepackage{upgreek}

\begin{document}


\title{Controlling the optical properties of transparent media by mixing
    active and passive resonances}

\author{Vikas Anant}
\author{Ayman F.\ Abouraddy}
\author{Karl K.\ Berggren}
\email[Electronic mail:]{berggren@mit.edu}
\affiliation{Research Laboratory of Electronics, Massachusetts Institute of Technology, Cambridge, Massachusetts 02139}



\begin{abstract}
Novel optical phenomena, including electromagnetically induced transparency, slow light, superluminal light propagation, have recently been demonstrated in diverse physical implementations.  These phenomena are challenging to realize in practical systems because they require quantum coherence as well as careful preparation and control of prescribed quantum states.  Here we present a unified approach to engineering optical materials that exhibit these phenomena by using mixtures of active and passive optical materials at frequencies near their resonances.  Our approach does not depend on quantum coherence and can realize large and small (much less than 1) indices of refraction and negative permittivity (\mbox{$\epsilon<0$}), normal and anomalous dispersion, all while maintaining transparency.
\end{abstract}

\pacs{need to put in OCIS codes}

\maketitle

\section{Introduction}

In recent years, the accepted wisdom regarding limitations of the optical properties of materials has been challenged.  Electromagnetically induced transparency (EIT) can render an otherwise opaque medium transparent to electromagnetic radiation using quantum interference \cite{eit}. Light can be sped up to superluminal group velocities, slowed down, or even stopped \cite{hau,stopped-light,stopped-light-1,stopped-light-2,bigelow}. While early demonstrations relied on systems exhibiting quantum coherence, consequently reducing their practicality, these desirable optical properties have also been proposed and achieved by schemes that either do not require quantum coherence or rely only on classical optical phenomena \cite{song,okawachi,howell,wang,wise,scully-pra}.

Many applications require low optical losses or, ideally, optical transparency.  However, achieving optical transparency in schemes such as those in Refs.\ \cite{song,okawachi,howell,wang,wise,scully-pra} is problematic because these schemes rely on the optical properties of resonant media.  On its own, a classical resonant medium can exhibit large or small susceptibility ($\chi$) or a large change in $\chi$ with respect to frequency \cite{scully-book}, albeit accompanied by strong absorption that masks these desirable optical phenomena. The same optical properties are achievable using an active resonance that results, for example, when population inversion is achieved in a laser gain material or a excited atomic gas \cite{loudon}.  But the strong amplification that accompanies an active resonance can be as problematic as strong absorption \cite{anant}.  Recent experiments have used gain or loss doublets as an attempt to mitigate these problems of loss and gain \cite{howell,wang}.  Two papers in the literature note that by mixing two materials, one passive and the other active, enhanced index \cite{scully-pra} and negative index \cite{ramakrishna} can be realized in truly transparent systems.  But these results were not applied more generally to demonstrate the wide variety of effects discussed above.

In this Letter we describe a conceptual framework that uses mixtures of active and passive resonances to realize a host of novel optical phenomena without quantum coherence and with transparency.  This framework is independent of the specific physical realization (nonlinear optics, atomic gas, solid-state, or combinations) allowing for new physical proposals.  We show how to control the amplitudes, widths, and resonant frequencies in order to reproduce in new physical systems a variety of \emph{lossless} optical phenomena including anomalous dispersion (\mbox{$\text{d}\chi/\text{d}\omega \gg 0$}, identical to EIT-related schemes \cite{eit,bigelow}), normal dispersion (\mbox{$\text{d}\chi/\text{d}\omega \ll 0$}), negative permittivity (\mbox{$\epsilon = 1+\chi <0$}), and large refractive index (\mbox{$n = (\chi^2+1)^{1/2} \gg 1$}).  Our main result is that these phenomena do not require quantum coherence to be realized and to occur with transparency.

We will first present the central concept in more detail.  We will then identify the specific optical phenomena that our scheme can realize, and finally present a possible physical system in which to implement this scheme.

\section{Description of central concept}

\begin{figure}
    \includegraphics[angle=90]{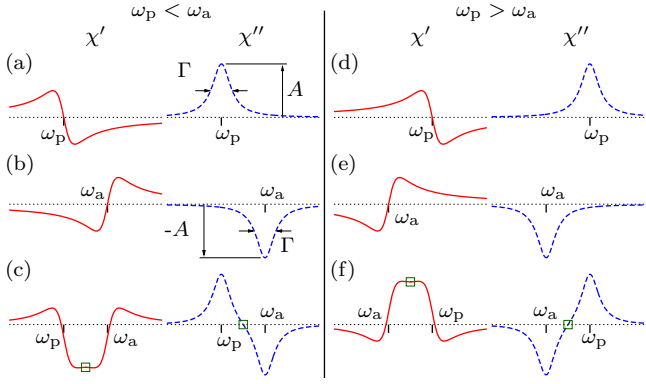}
    \caption{(color online). Plots of real ($\chi^\prime$) and imaginary ($\chi^{\prime\prime}$) components of susceptibility as a function of frequency $\omega$ for materials with \mbox{$\chi^\prime \ll 0$} or \mbox{$\chi^\prime \gg 0$}. If \mbox{$\omega_\text{a} > \omega_\text{p}$}, the susceptibilities of (a) passive and (b) active resonances can be combined to yield (c) a spectrum with \mbox{$\chi^\prime < 0$} and transparency. If \mbox{$\omega_\text{a} < \omega_\text{p}$}, the constituents shown in (d) and (e) can be mixed to realize (f) a spectrum with \mbox{$\chi^\prime>0$} and transparency.  Green squares in (c) and (f) indicate the susceptibility value at the frequency where transparency is achieved.}
    \label{fig.index}
    \end{figure}

Consider the susceptibility of a medium described by a generic (e.g.\ two-level atomic) resonance at frequency $\omega_0$:
    \begin{equation}
    \chi(\omega;A,\omega_0,\Gamma) = \frac{A (\Gamma / 2)}{(\omega_0-\omega)-i(\Gamma/2)}
    \label{eq.chi}
    \end{equation}
where $\omega$ is the probe frequency, $\Gamma$ is the linewidth, and $A$ is the amplitude; \mbox{$A > 0$} (\mbox{$A<0$}) for a passive (active) resonance.  Susceptibility, which relates the electric field to the polarization of the medium, is a complex quantity \mbox{$\chi = \chi^\prime + i \chi^{\prime\prime}$}, where \mbox{$\chi^{\prime\prime}>0$} (\mbox{$\chi^{\prime\prime}<0$}) corresponds to absorption (amplification).  Figure \ref{fig.index}(a) shows the real and imaginary parts of the susceptibility for a passive resonance \mbox{$\chi(A,\omega_\text{p},\Gamma)$}, while Fig.\ \ref{fig.index}(b) shows an active resonance \mbox{$\chi(-A, \omega_\text{a}, \Gamma)$}.  Note that desirable optical characteristics in these spectra (e.g.\ \mbox{$\chi^\prime>0$, $\chi^\prime<0$}, \mbox{$\text{d}\chi^\prime/\text{d}\omega>0$}, and  \mbox{$\text{d}\chi^\prime/\text{d}\omega<0$}) are masked by the strong absorption shown in Fig.\ \ref{fig.index}(a) and amplification in Fig.\ \ref{fig.index}(b).

Now consider an incoherent mixture of two optical resonances: one active and the other passive, accompanied by amplification and absorption, respectively. If the optical wavelength is longer than the length-scale of spatial inhomogeneity in this mixture, the effective susceptibility $\chi_\text{eff}$ of the mixture is given by
    \begin{equation}
    \chi_\text{eff} = \eta_\text{a} \chi_\text{a} + \eta_\text{p} \chi_\text{p}
    \label{eq.mixtures}
    \end{equation}
where $\eta_\text{a}$ and $\eta_\text{p}$ are the fractions of active (susceptibility $\chi_\text{a}$) and passive (susceptibility $\chi_\text{p}$) materials \cite{choy}.  By carefully choosing parameter values $A$, $\omega_0$, and $\Gamma$ to control the spectra of active and passive resonances, it is possible to realize either \mbox{$\chi_{\text{eff}}^\prime>0$}, \mbox{$\chi_{\text{eff}}^\prime<0$}, \mbox{$\text{d}\chi_{\text{eff}}^\prime/\text{d}\omega>0$}, or  \mbox{$\text{d}\chi_{\text{eff}}^\prime/\text{d}\omega<0$}, \emph{and} guarantee a transparency frequency where \mbox{$\chi_{\text{eff}}^{\prime\prime}=0$}.

In the next section, we outline how to choose parameter values to achieve either lossless negative-$\chi$ materials, high-index materials, superluminal light, or slow light, each accompanied by transparency.

\section{Novel optical phenomena achieved by mixing resonances}

Negative susceptibility (\mbox{$\chi^\prime<0$}) with transparency can be realized with a mixture of active and passive resonances.  The spectrum formed by a mixture of media with \mbox{$\chi(|A_\text{p}|, \omega_\text{p}, \Gamma_\text{a})$} and \mbox{$\chi(-|A_\text{a}|, \omega_\text{a}, \Gamma_\text{p})$}, where \mbox{$\omega_\text{a}>\omega_\text{p}$}, exhibits negative susceptibility at a zero-loss frequency. This spectrum is illustrated in Fig.\ \ref{fig.index}(c), where we have chosen \mbox{$\Gamma_\text{a} = \Gamma_\text{p}$} and \mbox{$\eta_\text{p} |A_\text{p}| = \eta_\text{a} |A_\text{a}|$}. These requirements need only be approximately satisfied in order for the effect to be observed. When \mbox{$0 < \chi^\prime(\omega_\text{t}) < -1$}, radiation at the transparency frequency $\omega_\text{t}$ will travel at superluminal group-velocities \cite{wang} in the resulting lossless low-index \mbox{$n < 1$} medium.  Media for which \mbox{$\chi^\prime(\omega_\text{t})<-1$} correspond to lossless negative-$\epsilon$ media \cite{pendry} and find utility in near-field lithography and microscopy \cite{fang}.
 
Alternatively, large $\chi^\prime$ accompanied by transparency can be obtained by mixing the same media but with \mbox{$\omega_\text{a} < \omega_\text{p}$}.  Such a mixture is potentially useful as an immersion fluid for nanometer-resolution immersion lithography and microscopy \cite{gil, rothschild}. Figure \ref{fig.index}(f) shows the spectrum resulting from choosing \mbox{$\Gamma_\text{a} = \Gamma_\text{p}$} and \mbox{$\eta_\text{p} |A_\text{p}| = \eta_\text{a} |A_\text{a}|$}. 

\begin{figure}
    \includegraphics[angle=90]{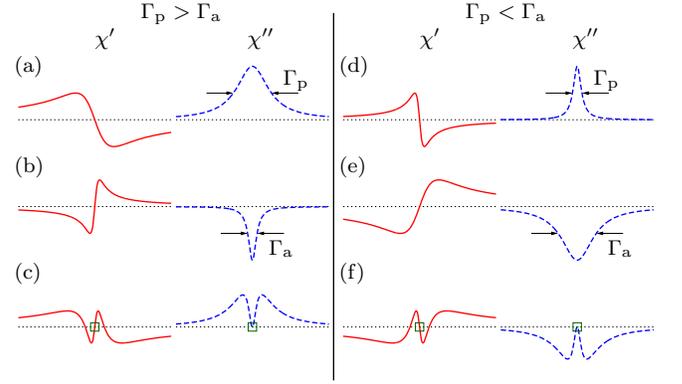}
    \caption{(color online). Real ($\chi^\prime$) and imaginary ($\chi^{\prime\prime}$) components of susceptibility as a function of frequency $\omega$ for materials with \mbox{$\text{d}\chi^\prime/\text{d}\omega \gg 0$} or \mbox{$\text{d}\chi^\prime/\text{d}\omega \ll 0$} accompanied by transparency (\mbox{$\chi^{\prime\prime}=0$}).  A spectrum with \mbox{$\text{d}\chi/\text{d}\omega \gg 0$} and transparency is shown in (c), where materials with susceptibilities shown in (a) and (b) are mixed.  The active resonance linewidth $\Gamma_\text{a}$ is narrower than the passive resonance linewidth $\Gamma_\text{p}$.  Alternatively,  if \mbox{$\Gamma_\text{p}<\Gamma_\text{a}$} as shown in (d) and (e), then the spectrum shown in (f) can be obtained where \mbox{$\text{d}\chi/\text{d}\omega \ll 0$} with transparency.  Green squares in (c) and (f) indicate the susceptibility value at the frequency where transparency is achieved.}
    \label{fig.slope}
    \end{figure}

Mixtures of resonances can also be used to realize ultraslow light (\mbox{$\text{d}\chi^\prime/\text{d}\omega \gg 0$}) and superluminal light (\mbox{$\text{d}\chi^\prime/\text{d}\omega \ll 0$}) with transparency.    Figure \ref{fig.slope}(c) shows a spectrum containing \mbox{$\text{d}\chi^\prime/\text{d}\omega \gg 0$} with transparency that was constructed by mixing passive (Fig. \ref{fig.slope}(a)) and active (Fig. \ref{fig.slope}(b)) media.  Creating such a spectrum would normally require quantum interference on a multi-level atomic system \cite{eit,hau,bigelow}.  However, in this case the spectrum was obtained by mixing equal proportions of media with \mbox{$\chi(|A|, \omega_0, \Gamma_\text{p})$}  and \mbox{$\chi(-|A|, \omega_0, \Gamma_\text{a})$} and choosing \mbox{$\Gamma_\text{a} < \Gamma_\text{p}$}.  An alternative choice for the relative magnitudes of active and passive resonance linewidths shown in Figs.\ \ref{fig.slope}(a-c) can result in a spectrum containing \mbox{$\text{d}\chi^\prime/\text{d}\omega \ll 0$} with transparency.  Such a spectrum is conducive to achieving superluminal group velocity and is shown in Figs.\ \ref{fig.slope}(d-f), where \mbox{$\Gamma_\text{a} > \Gamma_\text{p}$}.

\section{Proposed implementation}

\begin{figure}
    \includegraphics[angle=90]{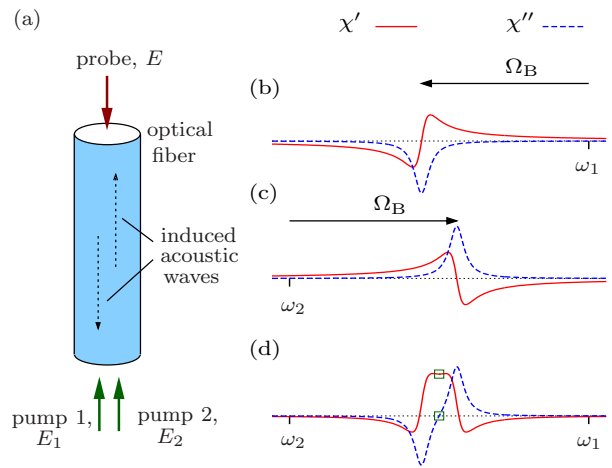}
    \caption{(color online). A schematic of a proposed physical realization of medium with either high or low index with transparency is shown in (a).  An active resonance, shown in (b), and a passive resonance, shown in (c), are generated using stimulated Brillouin scattering. By choosing pump frequencies that are separated by slightly less than $2\Omega_\text{B}$, one can create the probe spectrum shown in (d).  The spectrum shown in Fig.\ \ref{fig.index}(c) can be obtained by separating pump frequencies by slightly more than $2\Omega_\text{B}$. Green squares in (d) indicate the susceptibility value at the frequency where transparency is achieved.}
    \label{fig.new-sbs-fiber-and-resonances}
    \end{figure}

A number of possible implementations of this scheme are realizable.  For example, one could mix resonances from optical transitions of two different atoms in a solid-state system, e.g.\ two rare-earth dopants in a glass host.  Alternatively, one could mix resonances resulting from nonlinear optical phenomena such as stimulated Brillouin scattering (SBS) or stimulated Raman scattering (SRS) in the same material.  One can even combine an atomic and nonlinear-optical resonance, e.g. an absorption profile from dopant atoms and SBS.  In fact, these schemes can be realized in \emph{any} material where Eq.\ \ref{eq.mixtures} holds and where collision and broadening effects are minimal, regardless of the physical origin of the resonances.  We will now describe a specific example implementation where two SBS resonances are mixed to speed up or slow down an optical pulse.  Our implementation is shown schematically in Fig.\ \ref{fig.new-sbs-fiber-and-resonances}.

Stimulated Brillouin scattering (SBS) is a nonlinear optical phenomenon where energy transfer between oppositely-propagating optical waves is mediated by an acoustic wave with frequency $\Omega_\text{B}$ \cite{boyd}. An optical pump at frequency $\omega_\text{pump}$, via SBS, creates both an active resonance at \mbox{$\omega_\text{pump} - \Omega_\text{B}$} and a passive resonance at \mbox{$\omega_\text{pump} + \Omega_\text{B}$} that can interact with a counter-propagating probe optical wave.  Consider a medium in which two co-propagating pumps, with fields $E_1$ and $E_2$, and one counter-propagating probe with field $E$ have frequencies \mbox{$\omega_1<\omega<\omega_2$} and corresponding wavevectors \mbox{$k_1<k<k_2$}, and have fields given by
    \begin{subequations}\label{eq.E}
    \begin{eqnarray}
    E_m(z,t)&=&A_m(z,t) e^{i(k_m z-\omega_m t)}+\text{c.c.}, m=1,2\\
    E(z,t)&=&A(z,t) e^{i(-k z-\omega t)}+\text{c.c.}    
    \end{eqnarray}
    \end{subequations}
where $A_1$, $A_2$, and $A$ are slow-varying functions in time and space.  Two counter-propagating acoustic waves are generated via electrostriction at the difference frequencies \mbox{$\Omega_1=\omega-\omega_1$} and \mbox{$\Omega_2=\omega-\omega_2$} with wave vectors \mbox{$q_1=k+k_1$} and \mbox{$q_2=k+k_2$}.

Ignoring terms of the order \mbox{$(q_2-q_1)^2$} in comparison to $q_1^2$ and $q_2^2$ since \mbox{$q_1 \approx q_2$} \cite{boyd}, the resulting polarization for the probe wave is expressed in terms of active and passive resonances as
    \begin{equation}
    \bar{P} = \big[\chi_\text{bg} + \chi(A_\text{p},\omega_\text{p},\Gamma_\text{B})+\chi(-A_\text{a},\omega_\text{a},\Gamma_\text{B})\big]\bar{E},
    \end{equation}
where \mbox{$\omega_\text{p}=\omega_1+\Omega_\text{B}$}, \mbox{$\omega_\text{a}=\omega_2-\Omega_\text{B}$},  $\chi_\text{bg}$ is the linear background susceptibility, and $\Gamma_\text{B}$ is the Brillouin linewidth.  In this equation, \mbox{$\chi(A_\text{p},\omega_\text{p},\Gamma_\text{B})$} is the passive resonance that results via anti-Stokes SBS from the pump field at $\omega_1$, while \mbox{$\chi(-A_\text{a},\omega_\text{a},\Gamma_\text{B})$} is the active resonance that results via Stokes SBS from the pump field at $\omega_2$.  Parameters $A_\text{p}$ and $A_\text{a}$ are given by \mbox{$A_\text{p} =Sq_1|A_1|^2$} and \mbox{$A_\text{a} =Sq_2|A_2|^2$}, where $S$ is a constant encompassing the material parameters \cite{boyd}.  The resulting spectra of this mixture of active and passive resonances where \mbox{$A_\text{p}=A_\text{a}$} are shown in Figs.\ \ref{fig.index}(c) and (f) for \mbox{$\omega_\text{a}>\omega_\text{p}$} and \mbox{$\omega_\text{a}<\omega_\text{p}$}, respectively.
At the transparency point, the probe will experience \mbox{$\chi^\prime<\chi^\prime_\text{bg}$} (Fig.\ \ref{fig.index}(c)) and \mbox{$\chi^\prime>\chi^\prime_\text{bg}$} (Fig.\ \ref{fig.index}(f)).  This modification of $\chi^\prime_\text{bg}$ can be used to delay or speed up an optical pulse.  If we assume \mbox{$A_\text{p}=A_\text{a}$}, the maximum delay or speed-up of an optical pulse occurs when the resonances are spaced by $\Gamma_\text{B}$ (i.e. \mbox{$|\omega_\text{a}-\omega_\text{p}|=\Gamma_\text{B}$}). Calculations show that using a 1-km-long optical fiber made with Lucite \cite{pohl} at \mbox{$\lambda =$ 1.55~$\upmu$m}, \mbox{100~mW} pump power, and fiber core diameter of \mbox{1~$\upmu$m$^2$}, an optical pulse can be delayed or sped up by \mbox{85~ns}.  An experiment that demonstrates this delay, and hence, change in index with neither gain nor loss, will need a setup similar to one used for other slow- and fast-light experiments involving SBS \cite{song,okawachi} (in these experiments, the pulse was amplified/attenuated while being delayed/sped up, while in our case, the pulse amplitude is unchanged).

\section{Conclusion}

We have prescribed a framework that presents a simple approach to refractive index engineering.  Our framework does not require quantum coherence to exhibit \mbox{$\chi^\prime>0$}, \mbox{$\chi^\prime<0$}, \mbox{$\text{d}\chi^\prime/\text{d}\omega>0$}, or  \mbox{$\text{d}\chi^\prime/\text{d}\omega<0$} with transparency.  Implementations of our framework could permit application of novel optical methods to next-generation immersion optical lithography systems, near-field microscopy, and optical pulse storage.

\bibliographystyle{osajnl}
\bibliography{mixtures-of-materials}

\end{document}